\documentclass[a4paper,11pt]{article}
\usepackage{pos}
\usepackage{amsmath}
\usepackage{natbib}
\usepackage{hyperref}
\usepackage[nameinlink]{cleveref}
\DeclareMathOperator{\Tr}{Tr}

\title{Density profiles and correlations of harmonically trapped ultracold fermions via complex Langevin}
\ShortTitle{Harmonically trapped ultracold fermions via complex Langevin}

\author[a]{Felipe Attanasio}
\author*[a]{Marc Bauer}
\author[a,b]{Jan M. Pawlowski}

\affiliation[a]{Institute for Theoretical Physics, Universit\"{a}t Heidelberg, Philosophenweg 16, D-69120, Germany}
\affiliation[b]{ExtreMe Matter Institute EMMI, GSI, Planckstr. 1, D-64291 Darmstadt, Germany}

\emailAdd{bauer\_m@thphys.uni-heidelberg.de}

\abstract{
	Standard lattice formulations of non-relativistic Fermi gases with two spin components suffer from a sign problem in the cases of repulsive contact interactions and attractive contact interactions with spin imbalance. We discuss the nature of this sign problem and the applicability of stochastic quantisation with complex Langevin evolution in both cases. For repulsive interactions, we find that the results converge,  using adaptive step size scaling and a Gaussian regulator to modify the lattice action. We present results on density profiles and correlations of a harmonically trapped system in one spatial dimension.

}

\FullConference{%
The 39th International Symposium on Lattice Field Theory,\\
8th-13th August, 2022,\\
Rheinische Friedrich-Wilhelms-Universität Bonn, Bonn, Germany
}

\begin{document}
\maketitle

\section{Introduction}

In the past two decades, the study of ultracold fermionic gases has received a large amount of attention from both theory and experiment. These systems offer a rich phase structure, at sufficiently small temperatures the system undergoes a second order phase transition from the atomic gas into a superfluid phase. In this phase one also observes a crossover at from a BCS regime at weak couplings to tightly bound atomic pairs forming composite bosons in a BEC regime. In the presence of population or mass imbalance one may encounter an exotic type of superconductivity, an  Fulde–Ferrell–Larkin–Ovchinnikov phase~\cite{PhysRev.135.A550, osti_4653415}, that is characterised by pairing at finite total momentum. State-of-the-art experiments allow for an unprecedented amount of control over the studied Hamiltonian by tuning the interaction via Feshbach resonances. Moreover, harmonic traps for the atomic gas are commonly used, even though various shapes for the confining potential can be tuned by now. 

In many physically interesting cases the systems exhibit strong correlations, and hence non-perturbative methods are required for sound theoretical predictions. The determinant quantum Monte Carlo (DQMC) approach~\cite{PhysRevD.24.2278} is one such method that works by trotterising the thermal direction while putting the spatial direction on a lattice. DQMC has been employed in various systems, ranging from one to three spatial dimensions, to study systems in various confining potentials~\cite{Berger_2015, Wolak_2012, Wolak_2011}. The method is severely obstructed by the fermion sign problem that occurs in systems with a  finite population or mass imbalance, or in systems with repulsive contact interactions. 

Stochastic quantisation with a complex Langevin (CL) evolution offers a possible avenue to solving the sign problem. It has been first proposed in the 1980s~\cite{PARISI1983393, 10.1007/978-3-7091-7651-1_8} with applications mainly in the realm of lattice simulations for QCD~\cite{sexty_simulating_2014,aarts_qcd_2016,sexty_calculating_2019,attanasio_qcd_2022}. More recently, the method has been applied to systems of non relativistic fermions and bosons~\cite{Heinen:2022eyh, PhysRevA.101.033617, Hayata_2015} in one dimension, both at zero~\cite{Rammelm_ller_2017,Rammelm_ller_2020} and finite temperature~\cite{Loheac_2017, Loheac_2018, PhysRevResearch.3.033180}, as well as the three dimensional unitary Fermi gas~\cite{Rammelm_ller_2018, Attanasio:2022mjd}. In the present contribution we aim at the evaluation of the applicability of this approach, as well as reporting on first results on one-dimensional systems in harmonic trapping potentials.

\section{Ultracold fermionic gas in one dimension with a harmonic potential}

The Hamiltonian of a two-component Fermi systems in a harmonic confining potential and contact interaction between the spin species in one spatial dimension is given by 
\begin{align}
	\hat{H} = \hat K +  \hat{V}_{\text{int}} + \hat{V}_{\text{ext}} \,, 
\label{eq:hatH}
\end{align}
with the kinetic term $\hat{K}$ and the interaction $ \hat{V}_{\text{int}}$, 
\begin{align}
\hat{K}= \sum_\sigma \int \text{d}x \, \hat{\psi}^\dagger_\sigma (x ) \frac{-\nabla^2}{2m} \hat{\psi}_\sigma (x ) \,,  \qquad \qquad 	\hat{V}_{\text{int}} =  -g \int \text{d}x \, \hat{n}_\uparrow (x) \hat{n}_\downarrow(x)\,, 
\label{eq:K+IntPot}
\end{align}
as well as the trapping potential  
\begin{align} 
\hat{V}_{\text{ext}} = \frac{1}{2} m \omega^2 \sum_\sigma \int \text{d}x \, x^2 \hat{n}_\sigma (x)  \,.
\label{eq:TrapPot}
\end{align}
In \labelcref{eq:TrapPot}, $\omega$ is the trapping frequency. In our lattice setup, we will use the dimensionless quantity $\lambda = \sqrt{\beta} g$ given the inverse temperature $\beta$, instead of the bare coupling $g$ in \labelcref{eq:K+IntPot}. We work in the grand canonical ensemble, where the partition function is 
\begin{equation}
	Z = \Tr{ \left [e^{-\beta(\hat{H}- \sum_\sigma \mu_\sigma \hat{N}_\sigma)}\right]}\,,
\end{equation}
with particle number operator $\hat{N} = \sum_\sigma \int_x \hat{n}_\sigma (x)$ and chemical potential $\mu_\sigma$ related to each spin species. The trap introduces a length scale to the system, which we denote as
\begin{equation}
	L_T = \left(  m \omega \right)^{-\frac{1}{2}}\,.
\end{equation}
For the simulations we require this length to be much larger than the lattice spacing and much smaller than the size of our system. 

\section{Complex Langevin}  

To make the problem amenable to field-space Monte Carlo simulations, the Hamiltonian is put on a spatial lattice while the time direction is trotterised. We employ a bounded but continuous auxiliary field transformation of the form
\begin{equation}
	e^{\lambda \Delta t \hat{\psi}_{\uparrow,x}^{\dagger} \hat{\psi}_{\uparrow,x}^{\vphantom{\dagger}} \hat{\psi}^\dagger_{\downarrow,x}\hat{\psi}_{\downarrow,x}^{\vphantom{\dagger}}} = \int_{-\pi}^\pi d \phi_{x,t} \prod_\sigma \left( 1 + C  \hat{\psi}_{\downarrow,x}^\dagger \hat{\psi}_{\uparrow,x}^{\vphantom{\dagger}} \sin{\phi} \right) \,,
\end{equation}
where we obtain an effective coupling of the fermionic to the bosonic degrees of freedom $C = \sqrt{2(e^{\Delta t \lambda} - 1)}$ and $\Delta t$ is the Trotter time step. This procedure leads to a factorisation in the spin species, yielding
\begin{equation}
	Z = \int \prod_{x,t} d \phi_{x,t} \, \det{\left( 1 + M_\uparrow[\pmb{\phi}]\right)} \det{\left( 1 + M_\downarrow[\pmb{\phi}]\right)} = \int \prod_{x,t} d \phi_{x,t} \, e^{-S[\pmb{\phi}]} 
	\label{eq::PartFunc}
\end{equation}
for the partition function with Fermi matrix $M_\sigma$. The action on the right hand side is the logarithm of the product of the determinants, to wit $S\left[ \phi \right] = - \log{\prod_\sigma \det{\left( 1 + M_\sigma[\pmb{\phi}]\right)}}$. The action remains real valued only as long as the product of determinants is real and positive. The lattice is periodic in contrast to the harmonic potential. Hence, we must ensure that the edges of the spatial lattice are sufficiently suppressed to avoid spillover effects.

In the case of an imbalance of the fermion species in either population or mass, or for repulsive contact interaction, the product of determinants is not necessarily real and positive. For a more thorough discussion of this, see section \Cref{sec:Applicability}. Recently, the application of complex Langevin simulations has been investigated for such systems, with applications to systems in one and three dimensions both imbalanced and repulsive~\cite{Attanasio:2021jkk, Loheac_2017, PhysRevD.96.094506, Loheac_2018, Rammelm_ller_2018, Shill_2018, PhysRevResearch.3.033180, Rammelm_ller_2020}. 

The main idea behind the complex Langevin approach is the complexification of field space, allowing the stochastic process to sample a real probability distribution on the extended space. The fields are evolved according to 
\begin{align}
	\partial_\tau \mathrm{Re}\left\{\phi\right\} &= \mathrm{Re} \left\{ \partial_\phi S\left[\pmb{\phi}\right] \right\} + \eta \,, 	\label{eq:langevin_real} \\[2ex]
	\partial_\tau \mathrm{Im}\left\{\phi\right\} &= \mathrm{Im}\left\{ \partial_\phi S\left[ \pmb{\phi} \right]  \right\}\,.
	\label{eq:langevin_complex}
\end{align}
In this case, $\tau$ represents a fictitious stochastic time. In \labelcref{eq:langevin_real,eq:langevin_complex}, all indices on the fields, as well as the Gaussian noise term $\eta$, are suppressed for clarity. In our simulations, noise is applied only to the real part of the evolution equation. While noise on the level of the imaginary part is possible, prior investigations have shown issues with stability and convergence~\cite{Aarts_2010, Aarts_2011}. 

Despite the non-trivial success of CL simulations in many systems, they can convergence towards wrong results. These issues are typically found in the presence of slow decay in distributions towards infinity or around singularities in the drift term. These singularities introduce boundaries in the integration domain, and trigger boundary terms~\cite{Nagata_2016, Scherzer_2019, PhysRevD.101.014501}. While good progress has been made towards remedying the boundary terms at infinity via corrections, comparatively little is known about ways to improve the situation in the presence of poles~\cite{Seiler_2020}.

Large excursions in field space can be avoided by introducing a Gaussian regulator term to the drift force ~\cite{Loheac_2017}. This additional force arranges for the trajectories not to wander far out back. In practice, when calculating the drift, the action is modified as $S \rightarrow S + \xi \pmb{\phi}^2$. We use a regulator with a strength of $\xi = 0.01$ for all Langevin simulations, including those with no sign problem. 

\section{Reliability considerations for complex Langevin simulations} 
\label{sec:Applicability}

In this section, we provide a short overview of the types of sign problems encountered in two component Fermi systems bosonised as indicated in Equation \ref{eq::PartFunc}. Then, they are discussed in the context of complex Langevin simulations. 

For attractive systems, we have a real Hubbard-Stratonovich coupling $C$, leading to a real Fermi matrix, and thus real determinants in the partition function. Indeed, for systems with balanced populations and masses and attractive interactions we have
\begin{equation}
	\lambda > 0,\, \mu_\uparrow = \mu_\downarrow:\qquad e^{-S} = \det(1+M_\downarrow)\det(1+M_\uparrow) = \det(1+M_\downarrow)^2 \in \mathbb{R}^{\geq 0}\,,
\end{equation}
resulting in a probability measure including the possibility of zeroes. This should be kept in mind, since in such situations ergodicity is a concern with drift-based methods like Langevin and hybrid Monte Carlo.
 
If we loosen the constraint on the balance of populations, the weight remains real but can also take negative values
\begin{equation}
 	\lambda > 0,\, \mu_\uparrow \neq \mu_\downarrow:\qquad  e^{-S} = \det(1+M_\downarrow)\det(1+M_\uparrow) \in \mathbb{R}\,,
 \end{equation}
yielding a genuine `sign problem' with positive and negative signs instead of a more general complex weight as encountered, for instance, in QCD. This non-positive-definite weight implies a generally complex valued action and a fermion sign problem. While results on spin imbalances in finite temperature systems appear promising, other models with  non-complex weight were shown to not be well suited for complex Langevin simulations. It was argued that a continuum stochastic process such as CL, in the limit of zero step size, cannot sample these systems due to the segregation or ergodicity problem stemming from singularities in the drift force~\cite{separation}. A different study~\cite{Aarts:2017vrv} found convergence towards quenched results instead, arguing this to be a discretization artifact, allowing the field to jump over zeroes in the weight.

In light of these  considerations, we defer further discussion of the imbalanced case to the upcoming publication. Indeed, preliminary results indicate the absence of a notable sign problem for wide parameter ranges altogether, similarly to what was found in untrapped one-dimensional systems and studies using standard DQMC methods~\cite{PhysRevD.98.054514, PhysRevResearch.3.033180}. 

Finally, for repulsive interactions with $\lambda < 0$, we have a purely imaginary HS coupling that results in a weight that is generally complex valued
\begin{equation}
		\lambda < 0:\qquad   e^{-S} = \det(1+M_\downarrow)\det(1+M_\uparrow) \in \mathbb{C}\,.
\end{equation}
In this case, spin or mass imbalance has no bearing on the type of sign problem encountered. Repulsive systems have been show to give reliable results for small ($\lambda=-0.85$) to moderate ($\lambda=-1.7$) couplings in one dimensional systems, but start to deviate from results obtained via worldline formulations for large couplings~\cite{PhysRevD.99.074511}.

\section{Attractive systems and pairing}

\begin{figure}
	\includegraphics[width=.5\textwidth]{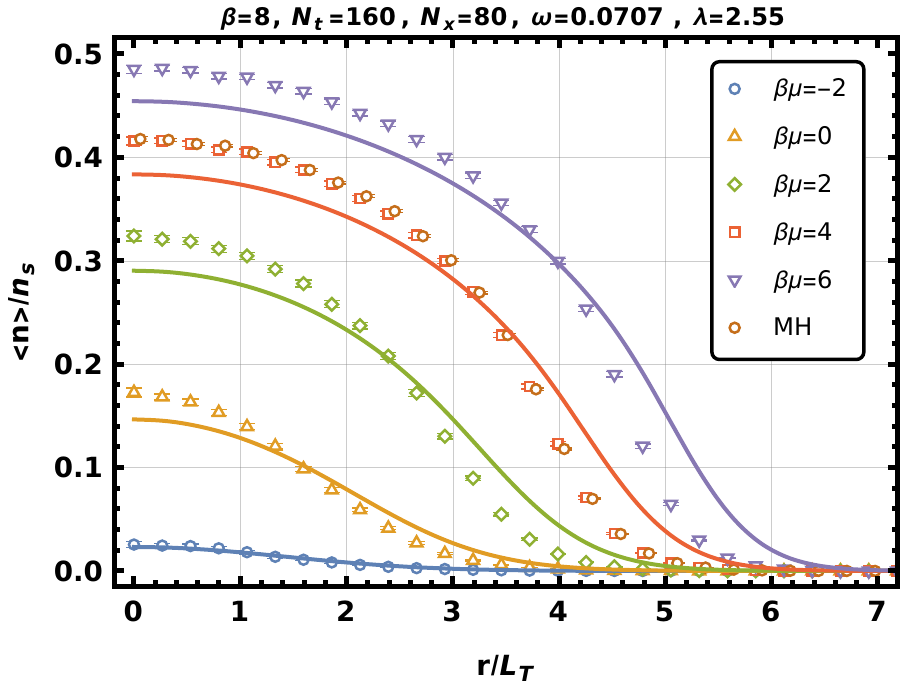}
	\includegraphics[width=.5\textwidth]{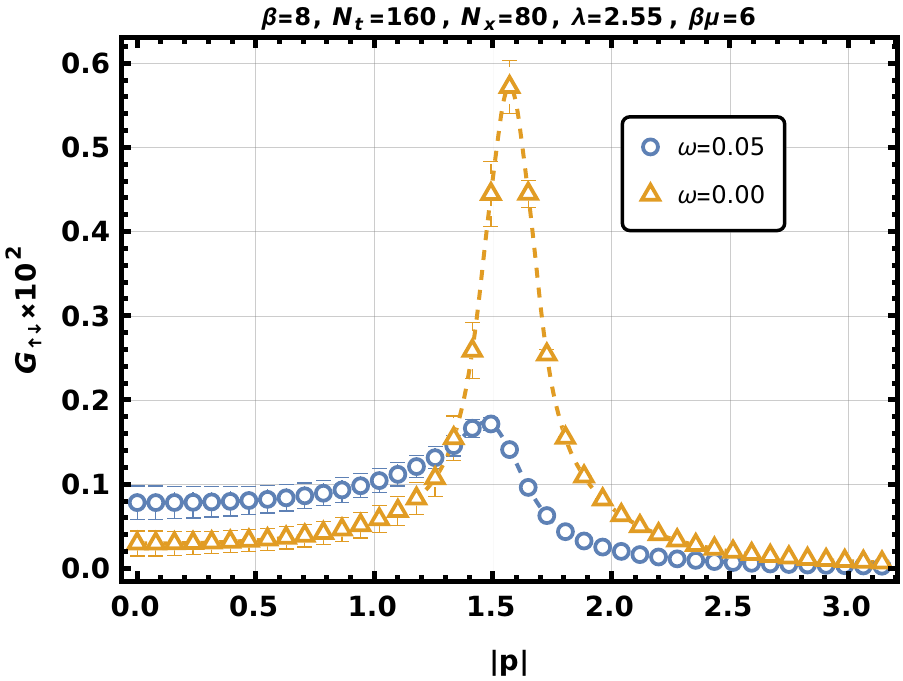}
	\caption{ LHS: Plot shows the density distribution for various chemical potentials and a comparison to Metropolis results, which are shifted for visibility. The position is given in length scales of the trap, while the density is normalized by its saturation value. Solid lines indicate the free case with matched average particle number. RHS: Connected density-density correlator of the trapped compared to the untrapped system at constant chemical potential with momenta in the Brillouin zone.}
	\label{fig::Attractive_balanced}
\end{figure}

We first test the standard Langevin approach in the case of an attractive system without a sign problem. In the following, $m=1$ in lattice units is taken for all cases. We use an inverse temperature of $\beta = 8$, with $N_t=160$ time slices, leading to a $\Delta t = 0.05$ Trotter time discretisation. The average Langevin step size is set to $\Delta \tau = 10^{-2}$ throughout, using an adaptive step size to avoid runaways~\cite{Aarts_2013}. We found this to be sufficient for our results to converge within the statistical error. We use a spatial lattice with $N_x = 80$ points, putting us well above both the length scale of the trap and the thermal wavelength. These scales are given by $L_t = 3.76$ and $\lambda_T = 7.1$ in lattice units. 

On the left-hand side of \Cref{fig::Attractive_balanced}, the density profiles are compared to their free counterparts for various chemical potentials. Here, we tuned the chemical in the free case such that we get the same average particle content as in the interacting one. We find an enhancement of the peaks in the center, which is expected for an attractive interaction. Because the average particle content is low, the effect is barely visible in the $\beta \mu =-2$ case. Additionally, we have compared the Langevin result for $\beta \mu =4$ to runs performed by Metropolis sampling with global change proposals and find the density profiles agree within error. Note that the Langevin result was sampled in the presence of a finite regulator as well as finite step size, without performing an extrapolation to zero. Hence, some deviations are expected given the large statistics. On the other hand, we can rule out ergodicity issues since the Metropolis never sees a sign change in the determinants, indicating the absence of boundaries in the relevant regions in configuration space.

In addition to the density profile, we have measured the connected density-density correlator, also called shot-noise in the experimental context, 
\begin{equation}
	G_{\uparrow,\downarrow}(k) = \langle \hat{n}_\uparrow (k) \hat{n}_\downarrow (-k) \rangle - \langle \hat{n}_\uparrow (k) \rangle \langle \hat{n}_\downarrow (-k) \rangle\,.
\end{equation}
The plot on the right side of \Cref{fig::Attractive_balanced} shows the shot-noise correlator for a system at fixed chemical potential with the trapping potential on and off. In the untrapped case, a distinct peak at the Fermi momentum of the system is visible, indicating the presence of BCS type pairing around the Fermi surface. In the presence of a harmonic trapping potential, the peak appears less distinct but does not fall off as steeply towards smaller momenta.

For a discussion of the systems in the presence of a finite population imbalance, we refer to the forthcoming publication.

\section{Density profiles for repulsive systems}

\begin{figure}
	\centering
	\includegraphics[width=.49\textwidth]{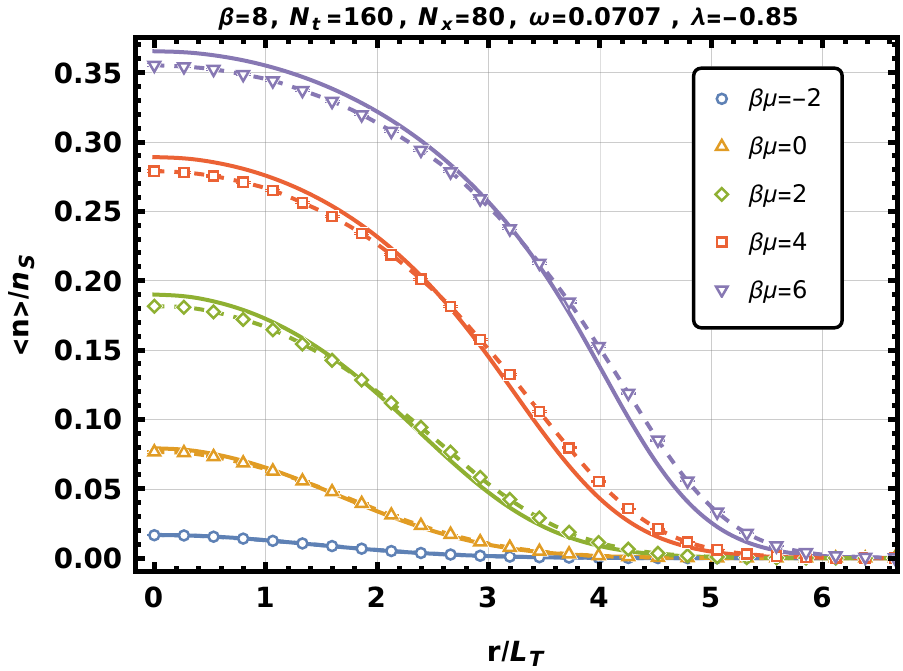}
	\includegraphics[width=.49\textwidth]{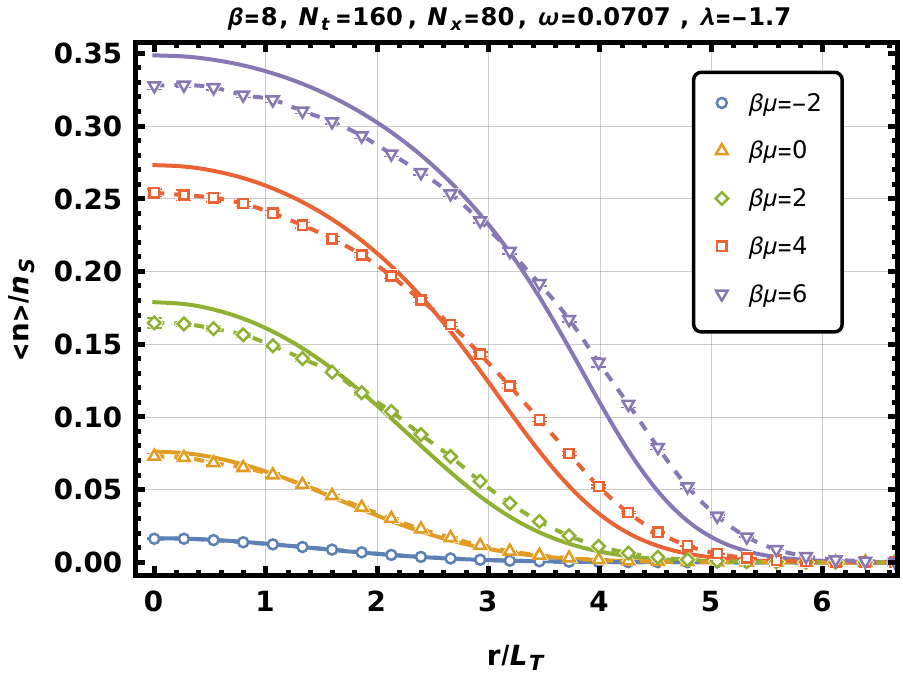}
	\caption{Density profiles for small (left) and moderate (right) repulsive interactions at various chemical potentials in the presence of a harmonic trapping potential. The dashed lines are spline fits to the data, while solid lines indicate the free case with matched average particle number}
	\label{fig::Repulsive_balanced}
\end{figure}

In repulsive systems, a sign problem is present, or more precisely, a complex action problem. This is the regime, where standard algorithms fail and we have to resort to complex Langevin simulations, or other approaches that potentially circumvent sign problems. Working again with a trotterisation step size of $\Delta t = 0.05$, an inverse temperature $\beta = 8$, a spatial lattice size $N_x=80$, and a trap length of $L_t=3.76$, results for the density profiles at small($\lambda = -0.85$) to moderate($\lambda = -1.7$) interaction strengths are depicted in \Cref{fig::Repulsive_balanced}. We see a similar effect to the attractive system, with the distribution being flattened and pushed out compared to its free, particle-content matched counterpart, with the effect increasing with stronger repulsion.

Since the stochastic process now explores the complex plane, we have to monitor the occurrence of potential boundary terms that may spoil the convergence towards the correct results. Indeed, as was found in previous studies, repulsive interactions induce some degree of slow decay in the field and drift distributions. However, while slow decay is an indicator of possible wrong convergence, there is no one-to-one correspondence. This was observed in systems at zero temperature for moderate interaction strengths, where complex Langevin results agree with those obtained by a formulation with hard-wall bosons~\cite{PhysRevD.99.074511, PhysRevD.96.094506}. Overall, further investigations are required for reaching a final verdict on the impact of boundary terms in these systems, especially on the possibility of correction terms.

\begin{figure}
	\includegraphics[width=.49\textwidth]{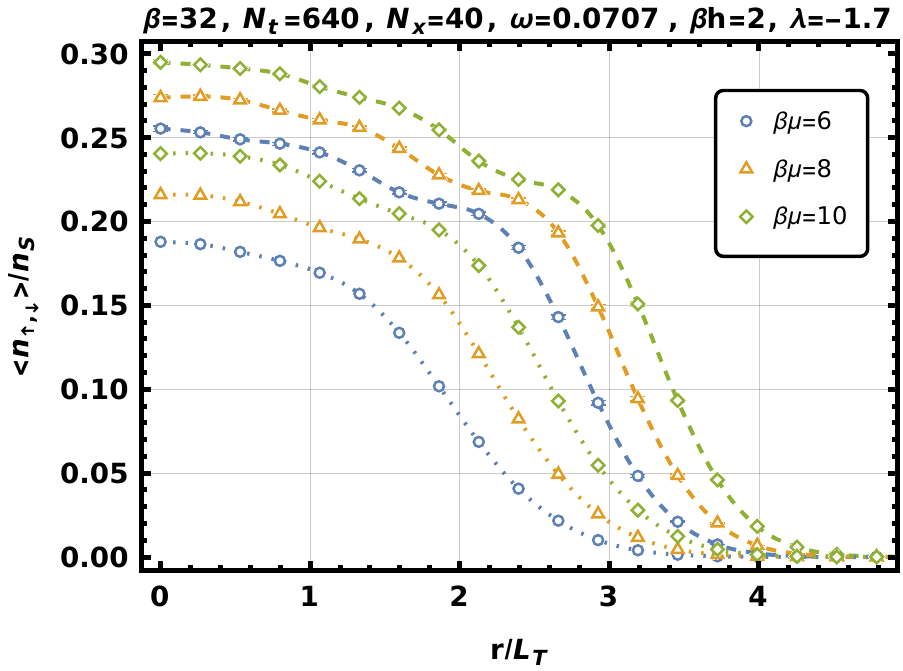}
	\includegraphics[width=.49\textwidth]{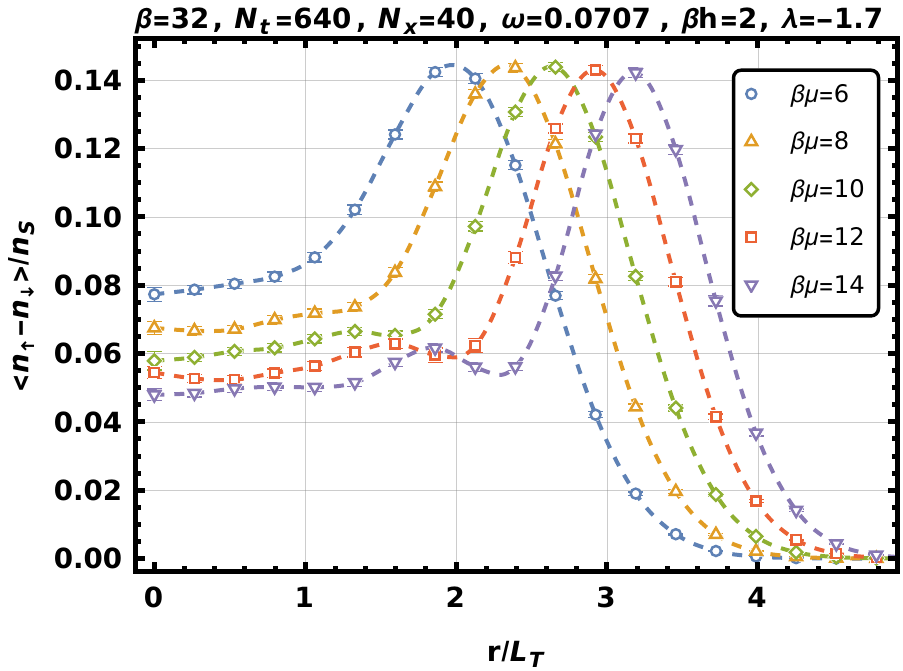}
	\caption{LHS: Density profile of spin imbalanced systems at various particle contents. Dotted lines are the minority, dashed lines the majority species. RHS: Local polarisations. The lines are spline fits to the data.}
	\label{fig::repulsive_imbalanced}
\end{figure}

Finally, the repulsive case can easily be extended to imbalanced mixtures of spins, since there is no qualitative change in the nature of the phase problem under investigation. We rewrite the chemical potentials using
\begin{equation}
	\mu = \frac{\mu_\uparrow + \mu_\downarrow}{2} \hspace{2cm} h = \frac{\mu_\uparrow - \mu_\downarrow}{2}\,.
\end{equation}
In \Cref{fig::repulsive_imbalanced} we show some results for a system with repulsive interaction $\lambda=-1.7$, an imbalance of $\beta h = 2$ and $\beta=32$, that is a significantly lower temperature than before, while keeping the Trotter step size constant. The spatial lattice size is $N_x = 40$, and the trap frequency is set to $\omega = 0.0707$. On the left side, the position space density profiles for various chemical potentials are shown. On the right, we plot the local polarisation, where we observe a maximum at finite radius, moving further out as the filling increases. We note that the bump in the majority profiles is not exclusively due to the repulsive interaction, but originates partly in the oscillations appearing in density profiles at lower temperatures. Some further caution is required here, since low temperatures can lead to a large separation of scales, and consequently may lead to a loss of precision in the calculation of the drift force and observables.

\section{Summary and outlook} 

We have discussed the application of complex Langevin simulation to trapped non-relativistic fermions in situations with attractive and repulsive couplings. The approach has been applied explicitly to the one-dimensional systems, and we have shown results for the density profiles and correlations. The systems with repulsive interactions exhibits a strong sign problem, and complex Langevin simulations may resolve the related issues. We indeed find, that density observables converge well in the given parameter regimes. Still, an incorrect convergence cannot be excluded due to sub-exponential decay in the distribution of drift and observables. In particular, the case of spin and mass imbalances necessitates further studies, since it presents a situation where complex Langevin simulations have been shown to struggle. As an extension, we are interested in the given system in the ground state, which is accessible via a projective approach, as well as the physically interesting projection of the present finite temperature results to fixed particle number. Another possible experimentally interesting application are systems in a rotating harmonic trap.

\section{Acknowledgments}

This work is funded by the Deutsche Forschungsgemeinschaft (DFG, German Research Foundation) under Germany's Excellence Strategy EXC 2181/1 - 390900948 (the Heidelberg STRUCTURES Excellence Cluster) and under the Collaborative Research Centre SFB 1225 (ISOQUANT). We also acknowledge support by the state of Baden-Württemberg through bwHPC.

\bibliography{references.bib}

\end{document}